# On the role of the Jeffreys'sheltering mechanism in the sustain of extreme water waves

Jean-Paul GIOVANANGELI, Christian KHARIF, Julien TOUBOUL

Institut de recherche sur les phénomènes hors équilibre, 49 rue F. Joliot-Curie, BP 146, 13384 Marseille cedex 13

**Abstract**. The effect of the wind on the sustain of extreme water waves is investigated experimentally and numerically. A series of experiments conducted in the Large Air-Sea Interactions Facility (LASIF) showed that a wind blowing over a strongly nonlinear short wave group due to the linear focusing of a modulated wave train may increase the life time of the extreme wave event. The expriments suggested that the air flow separation that occurs on the leeward side of the steep crests may sustain longer the maximum of modulation of the focusing-defocusing cycle. Based on a Boundary-Integral Equation Method and a pressure distribution over the steep crests given by the Jeffreys'sheltering theory, similar numerical simulations have confirmed the experimental results.

## 1. Introduction

Freak, rogue, or giant waves correspond to extreme waves surprisingly appearing on the sea surface (wave from nowhere) which produce serious damages to navigation and off shore industry. Over the two last decades dozens of super-carriers-cargo ships over 200m long have been lost at sea. Up to now there is no definitive consensus about a unique definition of freak wave. Sometimes, the definition of the freak waves includes that such waves are too high, too asymmetric and too steep. More popular now is the amplitude criterion of a freak wave: its height should exceed twice the significant height. There is a number of physical mechanisms producing the occurrence of freak waves. Extreme wave events can be due to refraction (presence of variable currents or bottom topography), dispersion (frequency modulation), wave instability (Benjamin-Feir instability), soliton interactions, etc. that may focus the wave energy into a small area. For more details on these different mechanisms see the review on freak waves by Kharif and Pelinovsky [1].

The main objective of this study is to better understand the physics of the huge wave occurrence in the presence of wind. To our knowledge, the present experimental and numerical study is the first one to consider the direct effect of the wind on the rogue wave formation. Herein, we use the spatio-temporal focusing phenomenon to generate freak wave events i.e the dispersive feature of surface gravity waves where long waves propagate faster than short waves. If initially short wave packets are located in front of long wave packets having larger group velocities, then during the stage of evolution, long waves will overtake short waves. A large-amplitude wave can occur at some fixed time because of the superposition of all the waves merging at the same location. Afterward, the long waves will be in front of the short waves, and the amplitude of the wave train will decrease.

## 2 - Experimental Study

The experiments have been conducted in the Large Air-Sea Interactions Facility (LASIF) of IRPHE at Marseille Luminy (Figure 1). A complete description of the facility is given in Favre and Coantic [2]. The water surface elevations η(x,t) have been determined by means of three capacitive wave gauges of 0.3mm outer diameter. One of these three wave probes was located at the fixed fetch of 1m from the upstream beach. The two other ones were installed on a trolley in order to determine the water surface elevation at different fetches all along the facility. The typical sensitivity of the wave probes was of order 0.6Volts/cm.

The initial wave train was generated by means of a computer-controlled wave maker submerged under the upstream beach.

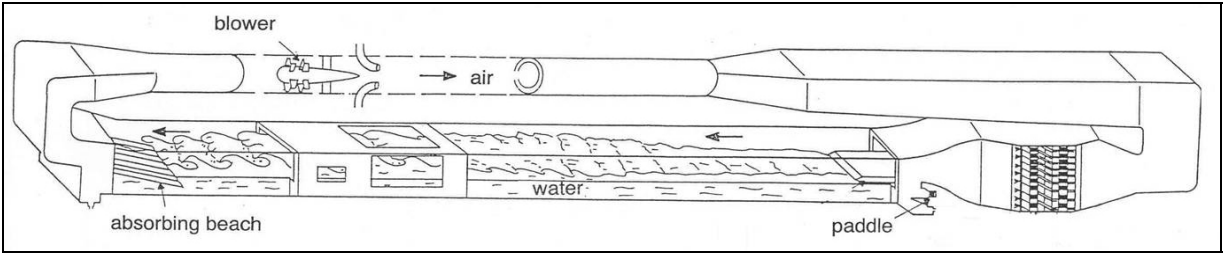

Figure 1 - A schematic view of Large Air-Sea Interactions Facility

Figure 2 presents the time series of $\eta(x,t)$ at different fetches x for $U = 0$ m/s, where U is the wind velocity. The initial wave train had a frequency varying from 1.3 Hz to 0.8 Hz in 10s. The repetability of the initial wave train with and without wind was verified. For clarity each time-serie are recursively increased by 10cm. As predicted by the linear theory of deep-water waves without wind, dispersion relation leads short waves to propagate slower than long waves. Then the waves focus at a precise fetch leading to the occurrence of a large amplitude freak wave (focusing). Downstream the focusing point, the amplitude of the group decreases rapidly (defocusing).

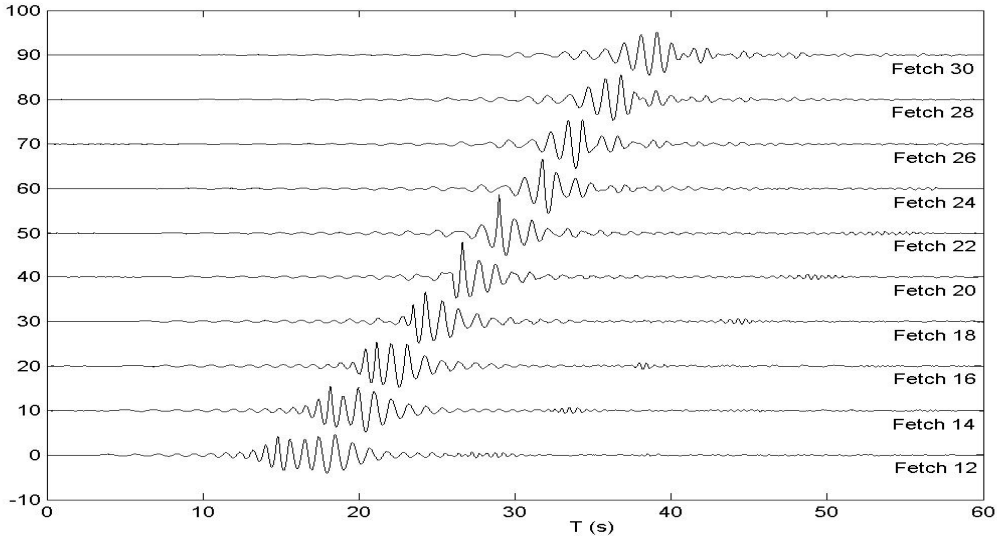

Figure 2 – Surface elevation time series (in cm) at different fetches (in m) for a wind speed U= 0 m/s.

Figure 3 presents the same time series of $\eta(x,t)$ at different fetches x, for a wind speed U= 6m/s and for the same initial wave train. Some evident differences appear between Figure 2 and Figure 3 in the space-time evolution of the focusing wave group. It appears that the amplitude of the largest wave inside the group seems to be maintained as the group propagates downstream.

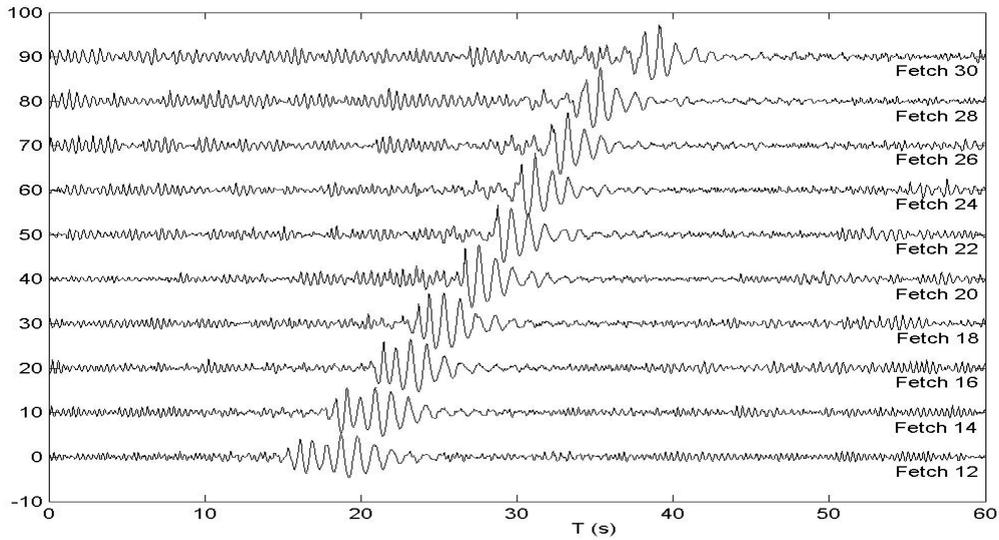

Figure 3 - Surface elevation time series (in cm) at different fetches (in m) for a wind speed U= 6 m/s.

For each value of the wind velocity, the amplification factor $A(x,U)$ of the group between fetch x and fetch 1m can be defined as:

$$A(x,U) = \frac{H_{max}(x,U)}{H_{ref}} \qquad (1)$$

where $H_{max}(x,U)$ is the maximal height between two consecutive crest and through in the group and $H_{ref}$ is the height of the quasi uniform wave train generated at fetch 1m. Here $H_{ref} = 6.13 cm$.

Figure 4 presents the evolution of $A(x,U)$ as a function of fetch x and for different values of the wind speed U equal to 0m/s, 4m/s and 6m/s. This figure shows that the effect of wind is twofold:(i) a weak increase of the amplification factor, and (ii) a downstream shift of the focal point are observed. Morerover, contrarely to the case without wind, an asymmetry appears between the focusing and the defocusing stage. This shows that the effect of wind on the freak wave mechanism is to increase its amplitude and its life time. It transforms the short group containing the freak wave into a long-lived short group, by delaying the defocusing stage.

To explain the obervations, Touboul et al [3] suggest that the Jeffreys'sheltering mechanism induced by an air flow separation process on the leeward face of the highest waves can be responsible of the significant increase of energy transfer from wind to the water wave group.

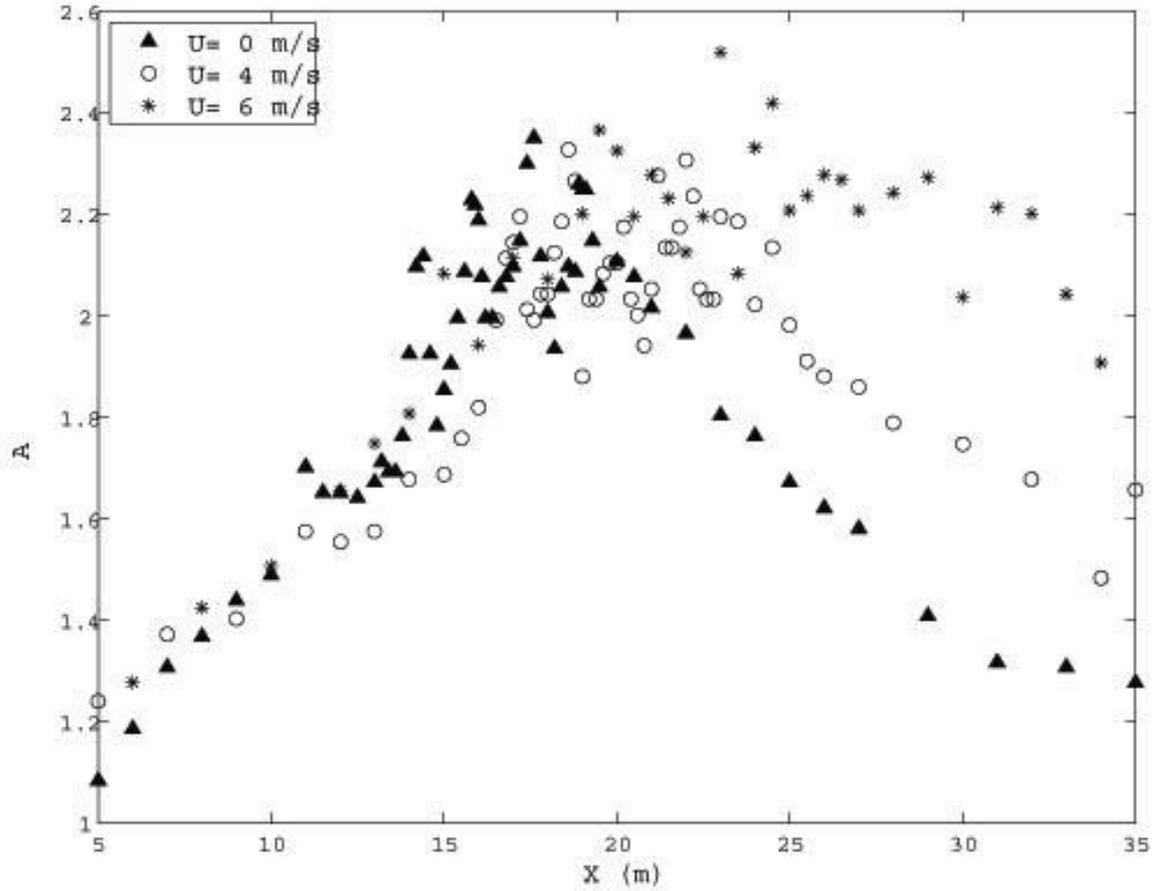

Figure 4 – Evolution of the experimental amplification factor $A(x,U)$ as a function of fetch x (in m) and for different wind speeds.

## 3. Numerical study

To model the direct effect of wind on water waves, two mechanisms can be considered: the Miles' mechanism, and the Jeffreys' sheltering mechanism. The Miles' mechanism, based on the concept of hydrodynamical instability, is a quasi laminar model of transfer of energy to a surface wave from a turbulent shear flow. The Jeffreys' sheltering mechanism assumes that the enhancement of the energy transfer between air flow and waves is due to an air flow separation occurring over very steep waves [4]. To see what is the dominant phenomenon prevailing in our experiments, one can calculate the ratio of the characteristic time scale of the transfer of energy from wind to waves for monochromatic wave trains of phase speed $c$.

$$\frac{T_{Jeffreys}}{T_{Miles}} = S \frac{\kappa^2}{\beta} \frac{(U-c)^2}{U_*^2} \qquad (2)$$

The sheltering coefficient, $S = 0.5$, is calculated from experimental data and $\kappa$ is the von Karman constant. The friction velocity is $U_* = U\sqrt{C_d}$ where the drag coefficient $C_d$ was determined experimentally equal to 0.004 and $U$ the mean wind velocity. The energy transfer parameter $\beta$ which is a function of the wave age can be computed from Figure 1 of [5].
The freak wave generated experimentally presents a peak at 1Hz. Thus, it is found that the Miles' characteristic time scale is roughly three times the Jeffreys' characteristic time scale.

Hence, we can consider that Jeffreys' mechanism is more relevant than the Miles' mechanism to describe the fast air-sea interaction process observed in the present experiments. Under the assumption of the air flow separation, Jeffreys [4] suggested that the atmospheric pressure at the interface is given by

$$p = \rho_a S(U-c)^2 \frac{\partial \eta}{\partial x} \tag{3}$$

where $\rho_a$ is atmospheric density and $\eta(x,t)$ is the interface elevation.

The problem is solved by assuming the fluid to be inviscid, incompressible, and the motion irrotational. Hence, the velocity field is given by $\vec{u} = \nabla \phi$ where the velocity potential $\phi(x,z,t)$ satisfies the Laplace's equation

$$\Delta \phi = 0 \tag{4}$$

Equation (4) is solved within a domain bounded by the fluid interface and solid boundaries of the numerical wave tank, subject to boundary conditions defined below.

The impermeability condition writes

$$\nabla \phi \cdot \vec{n} = \vec{V} \cdot \vec{n} \quad \text{on } \partial \Omega_B \tag{5}$$

where $\partial \Omega_B$ is the solid boundaries, $\vec{V}$ is the velocity of these solid boundaries, set equal to zero on the walls of the wave tank, and equal to the velocity of the paddle at any point of the wave maker and $\vec{n}$ is the unit normal vector to the boundaries.

A Lagrangian description of the interface is used

$$\frac{Dx}{Dt} = \frac{\partial \phi}{\partial x} \tag{6-a}$$

$$\frac{Dz}{Dt} = \frac{\partial \phi}{\partial z} \tag{6-b}$$

where $\frac{D}{Dt} = \frac{\partial}{\partial t} + \nabla \phi \cdot \nabla$.

The dynamic condition states that the pressure at the interface, $z = \eta(x,t)$, is equal to the atmospheric pressure

$$\frac{D\phi}{Dt} = \frac{1}{2}(\nabla \phi)^2 - gz - p \tag{7}$$

where $p(x,t)$ is given by equation (3). Air flow separation occurring above very steep waves, a critical wave slope is introduced beyond which equation (3) is applied, otherwise the pressure vanishes. This threshold is close to the maximal local slope of Stokes waves.

The equations are solved using a Boundary Integral Equation Method (BIEM) and a mixed Euler-Lagrange (MEL) time marching scheme. The Laplace's equation is solved by use of the Green's second identity. To avoid numerical instability the grid spacing $\Delta x$ and time increment $\Delta t$ have been chosen to satisfy the Courant relation derived from the linearized surface conditions. The discretization is such that the accuracy of the computions is $O(10^{-3})$

A 2D numerical wave tank simulating the Large Air-Sea Interaction Facility has been considered. The tank has a length of 40m, and a depth of 1m. A focusing wave train is generated by a piston wave maker, leading to the formation of a rogue wave followed by a defocusing stage. The free surface, and the solid walls (downstream wall, bottom and wave maker) are discretised by 1300, and 700 meshes respectively, uniformly distributed. The time integration is performed by using a RK4 scheme, with a constant time step $\Delta t = 10^{-2}$ s.

The numerical simulations have been run for several values of the critical wave slope, and wind velocity. Figure 5 shows the spatial evolution of the amplification factor computed numerically by using equation (1). These curves plotted for three values of the wind velocity, U=0m/s, 4m/s and 6m/s present the same behavior than those in Figure 4 and emphasize again the asymmetry found in the experiments. One can notice that the effect of the wind is not sufficient to increase the amplification of the freak wave. A very weak increase of the amplification factor is observed in presence of wind. The increase of the amplification due to the wind is significantly larger in the experiments. However, one can observe a significant asymmetry between the focusing and defocusing stages of the wave train. This asymmetry results in an increase of the life time of the freak wave event. Furthermore, a comparison between Figure 4 and Figure 5 shows that the numerical maxima of the amplification factor are larger than those obtained experimentally. This can be due in part to spilling breaking events which were observed in the experiments, resulting in dissipation of energy, and in saturation in the growth of the amplification factor. The comparison between experimental and numerical results should remain qualitative since the initial conditions used in the experiments and simulations are different. The fact that no downstream shifting of the focusing point is observed in the numerical simulations is due to the absence of current in the model.

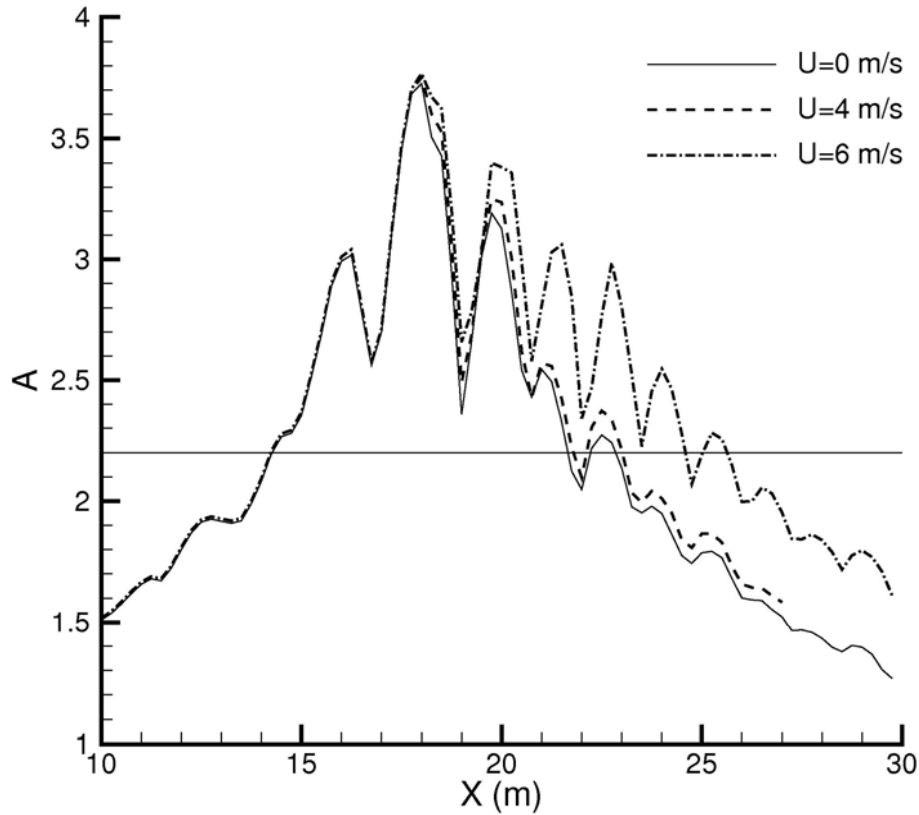

Figure 5 – Evolution of the numerical amplification factor $A(x,U)$ as a function of fetch (in m) and for different wind speeds.

**4. Conclusion**

The direct effect of the wind on a freak wave event generated by means of a dispersive spatio-temporal mechanism has been investigated experimentally and numerically. Both experiments and numerical simulations have shown that, in presence of wind an asymmetry in the amplification curve occurs. This asymmetry results in an enhancement of the life time of the freak wave event.